\documentclass[
 reprint,
 amsmath,amssymb,
 superscriptaddress,
]{revtex4-1}

\usepackage{graphicx}
\usepackage{dcolumn}
\usepackage{bm}
\usepackage{lineno}

\begin{document}


\thanks{A footnote to the article title}%
\title{Random laser action with coherent feedback via second-harmonic generation}

\author{Yanqi Qiao}
\affiliation{State Key Laboratory of Advanced Optical Communication Systems and Networks, Department of Physics and Astronomy, Shanghai Jiao Tong University, 800 Dongchuan Road, Shanghai 200240, China}
\affiliation{Key Laboratory for Laser plasmas (Ministry of Education), Collaborative Innovation Center of IFSA (CICIFSA), Shanghai Jiao Tong University, 800 Dongchuan Road, Shanghai 200240, China}

\author{Yuanlin Zheng}
\email{ylzheng@sjtu.edu.cn}
\affiliation{State Key Laboratory of Advanced Optical Communication Systems and Networks, Department of Physics and Astronomy, Shanghai Jiao Tong University, 800 Dongchuan Road, Shanghai 200240, China}
\affiliation{Key Laboratory for Laser plasmas (Ministry of Education), Collaborative Innovation Center of IFSA (CICIFSA), Shanghai Jiao Tong University, 800 Dongchuan Road, Shanghai 200240, China}

\author{Zengyan Cai}
\affiliation{State Key Laboratory of Advanced Optical Communication Systems and Networks, Department of Physics and Astronomy, Shanghai Jiao Tong University, 800 Dongchuan Road, Shanghai 200240, China}
\affiliation{Key Laboratory for Laser plasmas (Ministry of Education), Collaborative Innovation Center of IFSA (CICIFSA), Shanghai Jiao Tong University, 800 Dongchuan Road, Shanghai 200240, China}

\author{Xianfeng Chen}
\email{xfchen@sjtu.edu.cn}
\affiliation{State Key Laboratory of Advanced Optical Communication Systems and Networks, Department of Physics and Astronomy, Shanghai Jiao Tong University, 800 Dongchuan Road, Shanghai 200240, China}
\affiliation{Key Laboratory for Laser plasmas (Ministry of Education), Collaborative Innovation Center of IFSA (CICIFSA), Shanghai Jiao Tong University, 800 Dongchuan Road, Shanghai 200240, China}

\date{\today}

\begin{abstract}

The random laser action with coherent feedback by second-harmonic generation (SHG) was experimentally demonstrated in this paper. Compared with the conventional random laser action based on photoluminescence effect, which needs strong photoresponse in the active medium and has a fixed response waveband due to the inherent energy level structure of the material, this random SHG laser action indicates a possible confinement of the nonlinear signal with ring cavities and widens the response waveband due to the flexible frequency conversion in nonlinear process. The combination of coherent random laser and nonlinear optics will provide us another possible way to break phase-matching limitations, with fiber or feedback-based wavefront shaping method to transmit the emission signal directionally. This work suggests potential applications in band-tunable random laser, phase-matching-free nonlinear optics and even brings in new consideration about random nonlinear optics (RNO).

\end{abstract}

\maketitle

Anderson localization of photons and the interaction between light and random media have always attracted both theoretical and experimental physicists \cite{wiersma1997localization, makeev2003second, mel2004second}. Over the past decades, random laser action based on photoluminescence (PL) effect has provided us a new way to understanding this problem, where the 'ring cavity' formed by recurrent scattering was utilized to achieve light confinement and laser emission instead of standard optical cavities, such as Fabry-Perot (FP) resonators, whispering-gallery-mode resonators and photonic crystal cavities \cite{cao1999random, vanneste2001selective, mujumdar2004amplified, andreasen2011modes, wiersma2008physics, tulek2010naturally, vahala2003optical}. Recently, a random Raman laser was demonstrated in a three-dimensional system via stimulated Raman scattering (SRS) effect \cite{hokr2014bright}. However, this work indicates that no cavity confinement or coherent feedback is responsible for the bright emission. In general, previous random lasers need strong photoresponse in the active medium with a high efficiency of energy conversion via a four-level (or Raman-level) transition process \cite{hokr2014bright, jiang2000time}, and the response waveband is entirely limited by the inherent energy level structure of the material. Here, we experimentally demonstrate the random laser action with coherent feedback by SHG, which proves a possible confinement of the nonlinear signal with ring cavities. At the same time, this random SHG laser widens the response waveband due to the flexible frequency conversion in virtual state-based nonlinear processes, which indicates possible applications, for example, band-tunable random laser.

\begin{figure}[htb]
\centerline{
\includegraphics[width=8.5cm]{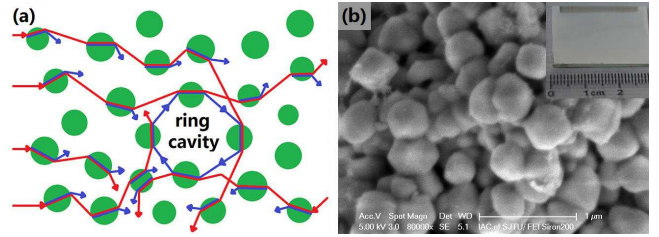}}
\caption{(a) The schematic of the FF laser (red light rays) and SH signals (blue light rays) inside the strong scattering medium (not exactly obey the refraction law). (b) A scanning electron microscopy image of the LN particles (scale bar:1 $\mu m$). Inset: a photograph of the ITO-LN powder film.}
\end{figure}

As is well known, for nonlinear optics, phase-matching (PM) or quasi-phase-matching (QPM) condition is always the limitation on its wide applications \cite{boyd2008nonlinear}. To break that, scientists have come up with many efficient methods, such as zero-index metamaterial \cite{suchowski2013phase} and random quasi-phase-matching (RQPM) \cite{baudrier2004random, fischer2006broadband, stivala2010random}. They utilized nonlinear materials with 'randomized domain' structure to achieve efficient nonlinear emission. Here, we propose that this random SHG laser action will also provide us a possible way to break these limitations, because our configuration is simpler and no PM or QPM condition is needed to achieve nonlinear laser emission. But one may notice that random lasers are usually extremely open systems. It is difficult to transmit stable and collimated nonlinear signals for further applications. Fortunately, a random fiber laser \cite{hu2012coherent} as well as a feedback-based wavefront shaping method \cite{vellekoop2007focusing, park2013subwavelength, park2014full} has been demonstrated recently, providing possible ways to transmitting the random signal directionally.

In this letter, we report the random SHG laser action with coherent feedback in superfine lithium niobate (LN) powder, with Fig. 1(a) showing the possible schematic of this process. All typical characteristics of random laser were demonstrated in our experiments. First of all, the backscattering cone was measured to describe the strong scattering inside the sample. In the SHG experiment, extremely sharp peaks were observed emerging in the broad SH spectra, indicating the possible laser modes. When considering the increasing intensity of one typical laser mode, we found a mutation point, indicating the threshold similar to that in conventional (random) laser action. Different SH laser spectra measured at different pump positions and emission angles were also observed to behave similarly. Finally, comparison experiments and a simplified theoretical modeling were carried out to further support our conclusions.

In our experimental setup, the grain size of the LN powder is very important because efficient SH emission as well as strong scattering needs to be satisfied. In conventional nonlinear optics, scientists usually utilize nonlinear crystals, which are relatively large and transparent for input and output light. Here, LN nanocrystal powder was prepared by a solid state reaction method, using niobium pentoxide ($Nb_2O_5$) and lithium acetate dihydrate ($C_2H_3O_2Li\cdot2H_2O$) as the reactants. The obtained particle has a maximum size of ~400 $nm$ [see Fig. 1(b)]. Then the powder was deposited onto an indium-tin oxide (ITO) coated glass substrate (thickness: ~1 $mm$) by electrophoresis method \cite{cao1999random, jeon1996electrophoretic}. The inset image in Fig. 1(b) shows a photograph of the ITO-LN powder film, indicating the sample size around $30\times30$ $mm^2$ and the average thickness around 100 $\mu m$.

\begin{figure}[htb]
\centerline{
\includegraphics[width=7.0cm]{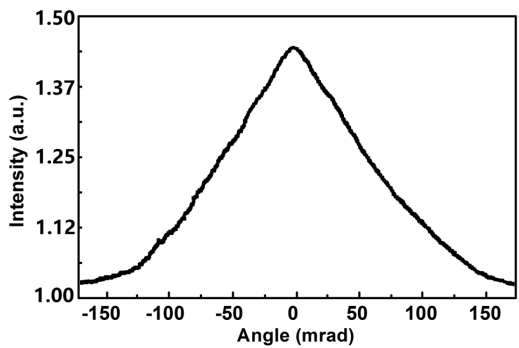}}
\caption{Coherent backscattering cone from a LN powder film at the wavelength of 405 $nm$. The average thickness is about 100 $\mu m$.}
\end{figure}

To characterize the light scattering properties in the LN powder film, we carried out the coherent backscattering (CBS) experiment \cite{van1985observation, wolf1985weak, de1996coherent} at the probe wavelength of 405 $nm$ (output of a violet diode laser). The measured backscattering cone is shown in Fig. 2. From the FWHM of the cone, the scattering mean free path was estimated to be $l\cong1.06\lambda$, according to $\theta\approx\lambda/(2\pi l)$ \cite{cao1999random, de1996coherent}. This value indicates an extremely strong scattering action inside the sample at 405 $nm$, which is close to the SH wavelength ($390\sim400$ $nm$) measured in the following experiment.

\begin{figure}[htb]
\centerline{
\includegraphics[width=7.0cm]{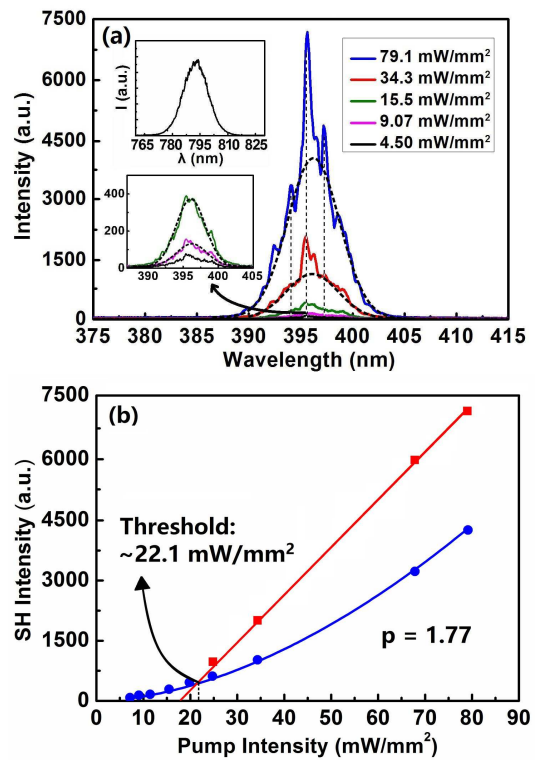}}
\caption{(a) Typical emission spectra of random SHG laser at various pump intensities. Upper inset: the spectrum of the pump centered at 793.5 $nm$. Lower inset: an enlarged view of the spectra at low pump intensities. (b) SH intensity versus the pump with a power fitting $I_{SH}\propto I_{FF}^p$ on the Gaussian background intensity shown in blue and the peak intensity beyond threshold at 395.5 $nm$ in red. The lasing threshold was determined to be 22.1 $mW/mm^2$.}
\end{figure}

In the broadband SH experiment, the light source was a Ti:Sapphire femtosecond regenerative amplifier system (50-fs duration, 1-kHz rep. rate). The center wavelength of the fundamental frequency (FF) laser was about 793.5 $nm$, as shown in the upper inset of Fig. 3(a). All spectra were measured using a high resolution optical spectrometer (SR500, Andor, resolution 0.1 $nm$). The normally pump position and the detecting angle $\theta$ were determined randomly for generality. When broadband FF laser entered the sample and scattered randomly, a second-order nonlinear process occurred in every pumped LN particle. These particles became SH sources, generating and radiating double frequency light in various directions [as shown in Fig. 1(a)]. Because of the strong scattering inside the powder, 'ring cavities' were formed by recurrent scattering, leading to the lasing action above certain pumping threshold. Figure 3(a) shows some typical spectra of random SHG laser at various pump intensities. When the pump was weak, there was no obvious laser peak because the gain did not exceed the loss, leading to a diffuse scattering spectrum of the emission (shown in green, purple and black). Some fluctuation emerging in these spectra should not be determined as the narrow peaks but the environmental noise, which would not increase as the increasing pump. Once the pump intensity increased beyond the threshold [22.1 $mW/mm^2$ as estimated from Fig. 3(b)], the gain exceeded the loss and a laser peak emerged in the spectrum (at 395.5 $nm$). As the pump intensity increased further, more sharp peaks were observed (at 394.0 $nm$ and 397.5 $nm$). It is worth mentioning that the bandwidth of these sharp peaks is around 1 $nm$, which is comparable to the results in the conventional random laser or standard cavity laser. (see more results in supplementary video S1)

To obviously determine the mutation point in Fig. 3(b), we used a Gaussian waveform to fit the background of the spectra in Fig. 3(a) and made a power fitting $I_{SH}\propto I_{FF}^p$ on the background intensity [shown in blue in Fig. 3(b)]. We arrived at $p=1.77$ which is not a perfect quadratic relation, probably due to the complex absorption and strong scattering inside the particles. Faez et al. also reported an obvious deviation ($p=1.87$) in a similar experiment except for the laser action \cite{faez2009experimental}.

The stability of the random SHG laser action was confirmed by repeated measurement while pumping the same position ($x$) and detecting at the same emission angle ($\theta$). After increasing the pump intensity and getting the emission spectra, we cut the pump power off and re-increased it repeatedly at different time intervals. The spectra were found almost the same, which demonstrated the stability and repeatability of the lasing action.

To further confirm the randomness of this process, we altered the pump position and detected at different angles, which were selected randomly for generality. Several typical emission spectra are shown in Fig. 4(a)-4(c) and multiple sharp peaks were observed. The emission spectrum in each condition has the same threshold characteristics, but the number and wavelength of these peaks are totally different. This result demonstrated the randomness of the scattering in the sample, which is also a typical feature of random laser action \cite{cao1999random}.

\begin{figure}[htb]
\centerline{
\includegraphics[width=6.5cm]{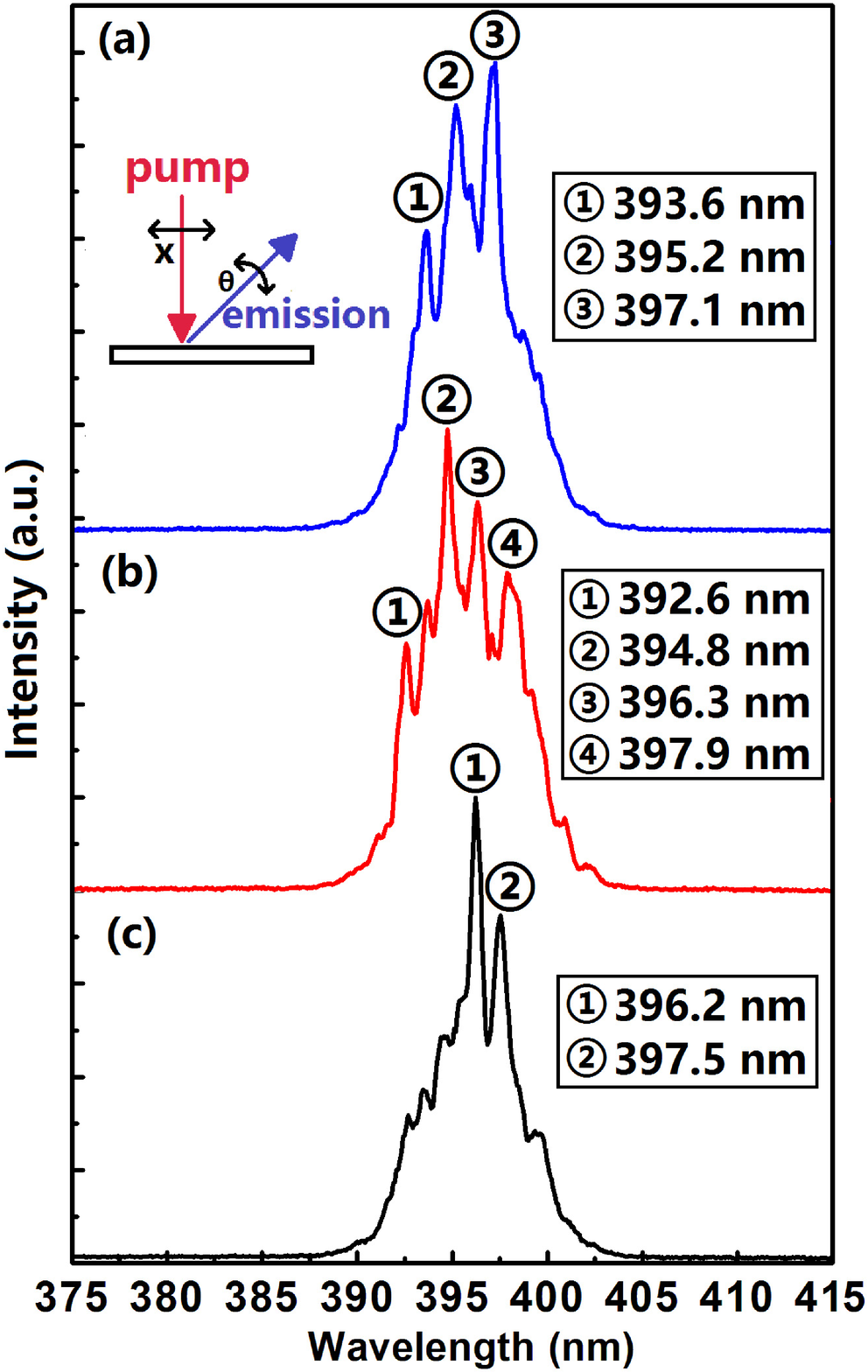}}
\caption{(a)-(c) Typical emission spectra of the random SHG laser at randomly selected pump positions ($x$) and emission angles ($\theta$).}
\end{figure}

 A comparison experiment was also carried out with LN powder loosely packed into a $10\times10$ $mm^2$ cuvette [shown in Fig. 5(a) inset], which means a weaker light scattering in the random system ($l\cong3.18\lambda$ from a CBS experiment, see the supplementary image S1). Figure 5(a) shows a typical series of SH spectra at various pump intensities set at similar levels (less than 100 $mW/mm^2$). As expected, the intensity of the Gauss-like spectrum increased uniformly with the pump, and no similar sharp peak was observed even with high intensity pump. Figure 5(b) shows the Gauss-spectra intensity as a function of the pump intensity and a power function was applied to fit the experimental data, arriving at $p=1.82$. No obvious mutation point or laser threshold could be determined based on these results. However, one may expect the laser emission in loosely packed LN powder with much higher pump intensity or additional light confinement, according to the theoretical and experimental results of the random lasers in weakly scattering systems \cite{hu2012coherent, vanneste2007lasing}. In fact, the relation of SH to FF intensity was maintained during our experiment even with increased pump power until the damage threshold of LN powder, which proves the absence of lasing action.

Furthermore, a same experiment was conducted on the LN-ITO system except for the frosted surface of the glass to rule out the possibility of an ITO formed FP cavity (see the supplementary image S2-S4). The LN powder film has no smooth surface, so this powder layer will not serve as an FP cavity. Besides, one particle serving as the cavity (something like micro-cavity or even nano-cavity) or two particles serving as the resonators (something like FP cavity) was considered as the special cases of ring cavities. Based on all the preconditions, a similar random SHG laser behavior was observed, which demonstrated that the observed lasing effect was truly caused by ring cavities instead of other possible conditions.

\begin{figure}[htb]
\centerline{
\includegraphics[width=7.0cm]{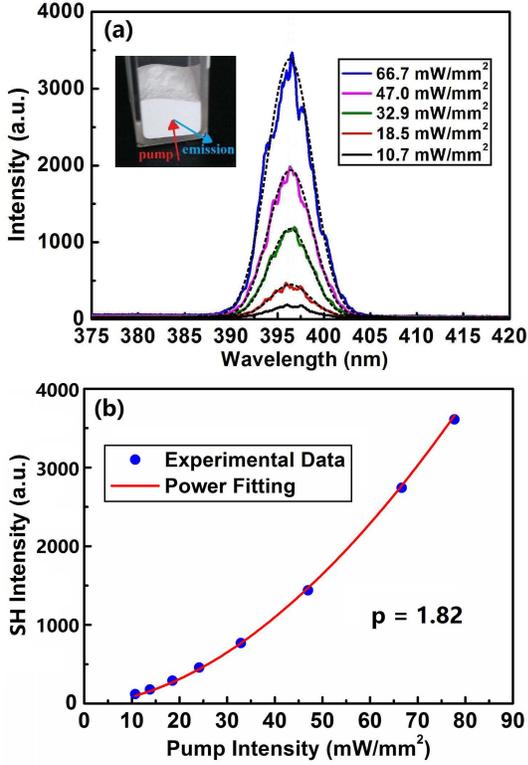}}
\caption{(a) Typical SH emission spectra of loosely packed LN powder at various pump intensity centered at 396.5 $nm$. Inset: the photograph of the sample. (b) The intensity of the Gaussian shaped spectra as a function of the pump intensity.}
\end{figure}

Up to now, there are not many theoretical demonstrations of coherent random laser action via nonlinear processes (without inherent energy state transition). The random Raman laser in 3D systems was a typical random nonlinear laser \cite{hokr2014bright}. However, the authors only presented the Monte Carlo simulations of the process without a detailed theoretical explanation. In fact, an early theoretical demonstration of resonant SH emission was proposed in 1D random structures \cite{centini2006resonant}, where a high conversion efficiency was predicted with four orders of magnitude higher than a bulk material and even one order of magnitude higher than an ideal phase matched slab of the same size. Based on the resonant SH emission in proposed 1D random laser structures and its amazingly high conversion efficiency, such mechanism to work in a 3D random SHG laser with coherent feedback can be naturally expected.

We begin with a simplified theoretical modeling and only focus on one effective cavity to present a possible mechanism of this lasing action. Due to multiple scattering, a closed loop may be formed for a single SH signal [blue light rays in Fig. 1(a)], but the loop is not resonant for a single pump light ray (in red) due to their different indices of refraction. Fortunately, multiple pump light rays scatter strongly in the random medium which may pass through the same route together with the SH signal inside the particles. Thus, it is reasonable to deal with the second-order process in one closed loop using the conclusions in RQPM theory \cite{baudrier2004random}, which can be represented as
\begin{equation}
I_{2\omega}=AI_\omega^2\cdot L_{eff},
\end{equation}
where $I_\omega$ is the pump intensity inside the loop, $L_{eff}$ is the effective gain length for enhancing the SH signal and $A$ is a constant. For a loss-free approximation, the SH enhancement in one loop walk can be represented as
\begin{equation}
I=I_0+AI_\omega^2\cdot L_{eff}=I_0(1+\dfrac{AL_{eff}}{I_0}I_\omega^2)\cong I_0e^{\frac{A}{I_0}I_\omega^2\cdot L_{eff}},
\end{equation}
where $I_0$ is an initial signal along the scattering loop and is considered to be independent of $I_\omega$. After introducing the loss term $I_0e^{-\alpha L}$ into the process, we get the final expression as
\begin{equation}
I\cong I_0e^{(\frac{A}{I_0}I_\omega^2-\alpha)\cdot L_{eff}},
\end{equation}
which is similar to the expression in traditional optical cavities. When the pump intensity $I_\omega$ increases, the gain may exceed the loss and $I>I_0$, indicating the occurrence of a laser action.

In our experimental design, LN particle was chosen for its large second-order susceptibility and high refractive index ($n\approx2.2$) for strong scattering. The average size of the particles is much smaller than the SH coherent length in LN crystal ($L_c$ is about several microns), which ensures the maximum emission of every single SH source. Moreover, similar results have been observed in other nonlinear materials [for example, cubic boron nitride (cBN) powder (see the supplementary image S5 and video S2)], using the same experimental configuration. Other nonlinear processes \cite{boyd2008nonlinear}, such as third-harmonic generation (THG), four-wave mixing (FWM), and coherent anti-Stokes Raman scattering (CARS), etc., are also expected to behave the same in random systems with proper configurations.

In conclusion, we have observed random SHG laser action with coherent feedback in superfine LN powder. The sharp peaks emerging in the broad SH spectra and the threshold characteristics demonstrate the lasing action directly. The repeatability and a sensitive spatial dependence with the pump position and emission angle were also demonstrated. Compared to conventional random laser action and nonlinear optics, this combination will widen the response waveband due to the flexible frequency conversion in nonlinear processes and provide us another possible way to break phase-matching limitations, with fiber or feedback-based wavefront shaping method to transmit the signal directionally. Furthermore, new considerations about RNO are also expected (For example, very recently, the nonlinear feedback mechanism is of great significance in noninvasive focusing and imaging through random media \cite{katz2011focusing, katz2014noninvasive, lai2015photoacoustically}.), which may be another interesting topic in the future.\\

This work is supported in part by the National Natural Science Foundation of China under Grant Nos. 61125503, 61235009 and 11421064, the Foundation for Development of Science and Technology of Shanghai under Grant No. 13JC1408300.


%

\end{document}